\pdfoutput=1
\documentclass[
twocolumn,
showpacs,showkeys,preprintnumbers,
amsmath,amssymb,
aps,
prl,
]{revtex4-1} 

\usepackage{graphicx,color}
\usepackage{dcolumn}
\usepackage{bm}

\begin{document}
\title{Sequencing proteins with transverse ionic transport in nanochannels}
\author{P. Boynton and M. Di Ventra}
\affiliation{Department of Physics, University of California, San Diego \\ La Jolla, CA 92093-0319, United States}
%
\begin{abstract}
{\it De novo} protein sequencing is essential for understanding cellular processes that govern the function of living organisms and all post-translational events and other sequence modifications that occur after a protein has been constructed from its corresponding DNA code. By obtaining the order of the amino acids that composes a given protein one can then determine both its secondary and tertiary structures through structure prediction, which is used to create models for protein aggregation diseases such as Alzheimer's Disease. Mass spectrometry is the current technique of choice for {\it de novo} sequencing. However, because some amino acids have the same mass the sequence cannot be completely determined in many cases. Here, we propose a new technique for {\it de novo} protein sequencing that involves translocating a polypeptide through a synthetic nanochannel and measuring the ionic current of each amino acid through an intersecting {\it perpendicular} nanochannel. To calculate the transverse ionic current blockaded by a given amino acid we use a Monte Carlo method along with Ramachandran plots to determine the available flow area, modified by the local density of ions obtained from molecular dynamics and the local flow velocity ratio derived from the Stokes equation. We find that the distribution of ionic currents for each of the 20 proteinogenic amino acids encoded by eukaryotic genes is statistically distinct, showing this technique's potential for {\it de novo} protein sequencing.
\end{abstract}
\maketitle
Living organisms depend on proteins to carry out the genetic code and perform many vital cellular tasks like metabolism \cite{nelson2008lehninger}. To understand how a protein works one must understand its structure. Proteins are special because of how versatile they are in binding to other molecules, and the structure of these binding sites often indicate the precise use of a protein.

The first step in understanding protein structure is knowing the sequence of a protein, meaning the order of the amino acids that compose it. There are 20 amino acids that are used as building blocks by eukaryotic genes to make proteins, all of which have the same chain of atoms as a backbone. What distinguishes each amino acid is its side chain, which can span from a single hydrogen in the case of glycine (GLY) to containing an indole functional group in the case of tryptophan \cite{nelson2008lehninger}. For a protein to function these amino acids fold up into secondary and tertiary structures that expose features like binding sites, which can be predicted based on the protein sequence. Ongoing research attempts to understand protein aggregation diseases such as Alzheimer's Disease \cite{kelley2008simulating} by performing simulations of structure formation, which would not be possible without the knowledge of the components of the peptides and proteins involved. In addition, protein sequences allow the synthesization of other proteins, which is necessary to compensate for diseases like Diabetes Type I in which the body does not produce the necessary peptide hormone insulin \cite{katsoyannis1966insulin,johnson1983human}.

The most common method for {\it de novo} protein or peptide sequencing (namely sequencing a protein for the first time) is mass spectrometry, a technique that involves fractionating the peptide into many smaller peptides and then obtaining the mass-to-charge ratio of each new peptide from the mass spectrometer. The problem with this technique is that fractionation is often carried out with gel electrophoresis, which is inherently slow \cite{schmitt2003capillary}. In addition, fractionation must be repeated many times to obtain small enough peptides so that one can discern the composite amino acids from just the total mass-to-charge ratio \cite{standing2003peptide}. Also, {\it de novo} sequencing is sometimes impossible with this technique since some amino acids have the same mass and charge ({\it e.g.}, leucine and isoleucine).

Edman degradation is another common method for {\it de novo} protein or peptide sequencing that utilizes repeated chemical washing and N-terminal cleaving to identify the sequence of amino acids one at a time \cite{edman1950method}. However, Edman degradation suffers from the same issue of fractionation as mass spectrometry since devices can only reliably sequence peptides up to about 30 amino acids \cite{laursen1971solid}. Nonetheless, the end result of identification via chromatography of each singled out chemically modified amino acid is reliable, albeit slow, but does require the use of many reagents.
\begin{figure*}[t!h]
\begin{center}
\centerline{\includegraphics[width=.7\textwidth]{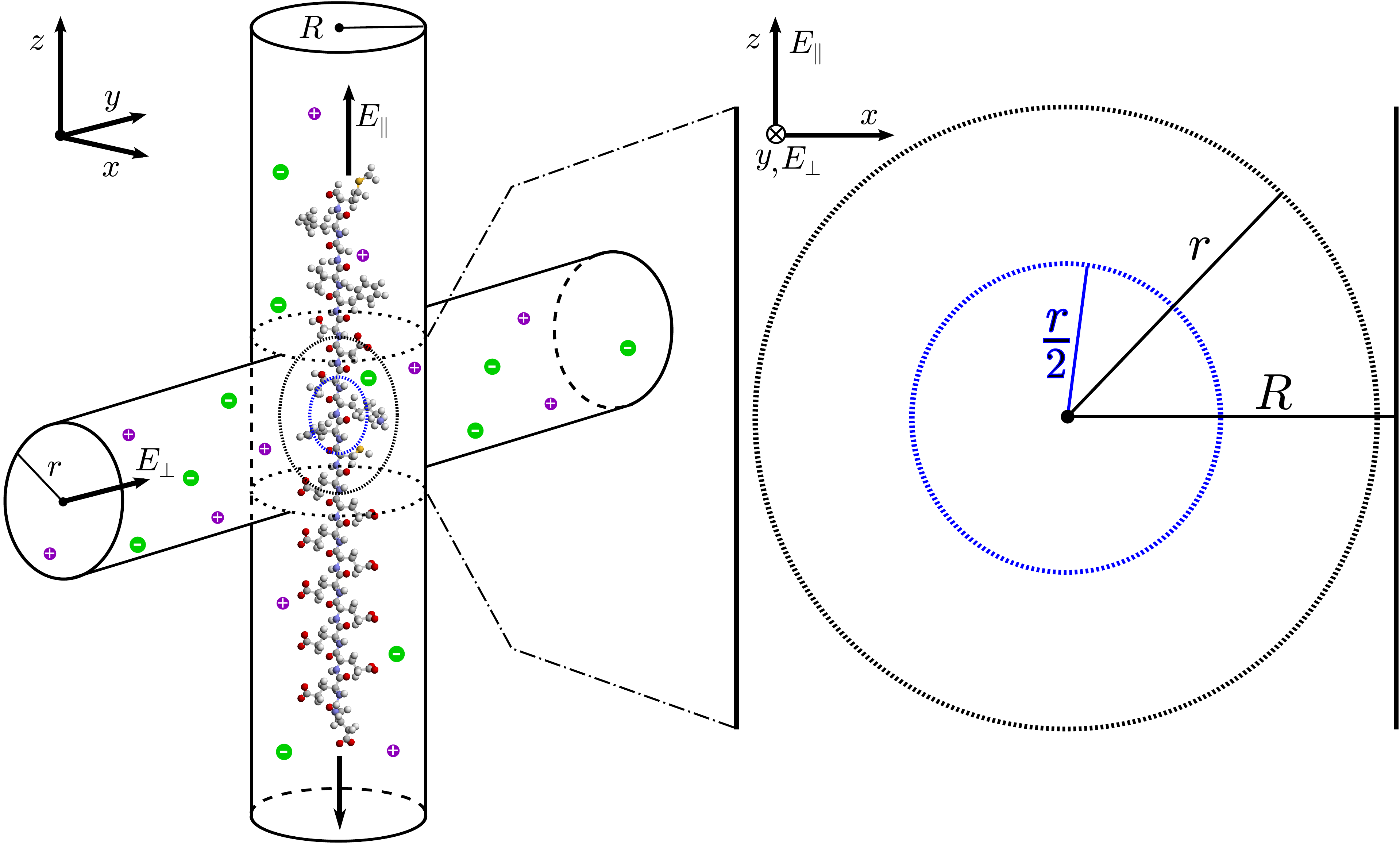}}
\caption{\label{fig:schematic}(Color online) A schematic of the transverse ionic transport sequencing method. Two nanochannels intersect: the vertical or longitudinal channel along $z$ with radius $R$ and the horizontal or transverse channel along $y$ with radius $r$. The polypeptide translocates along the longitudinal channel crossing the transverse channel that contains ions, purple ${\rm K}^+$ and green ${\rm Cl}^-$, that flow along the transverse channel due to an electric field, $E_{\perp}$, in the $+y$ direction. In this case the polypeptide consists of neurokinin A starting at the C-terminus at the top of the figure attached to one cysteine (CYS) followed by 10 glutamic acids (GLUs), a negatively charged amino acid, where the last GLU makes up the N-terminus (see section Sequencing Protocol for more on this structure). This negatively charged polypeptide is driven towards $-z$ by an electric field, $E_{\parallel}$, in the $+z$ direction. The dotted lines represent the top and bottom extremities of the intersection of the transverse channel, which are expanded to the right along with the thick dashed lines representing the area-limiting cross section (outer black line) and the Monte Carlo radial limit (inner blue line) that lie in the $xz$-plane. For visibility purposes the polypeptide is enlarged by a factor of 3 in both of its dimensions from the actual scale that we used in simulations while the ion radius is enlarged by a factor of 1.5.}
\end{center}
\end{figure*}

The advent of nanopore DNA sequencing \cite{Zwolak2008,branton2008potential} has brought several modern techniques to protein detection: longitudinal ionic transport \cite{talaga2009single,movileanu2009interrogating} and transverse electronic transport \cite{ohshiro2014detection}. In the case of ionic transport through a single nanopore, detection of the protein folding state is achieved experimentally and modeled with exclusion volumes by \cite{talaga2009single}. Of course, protein sequencing with such a technique is a more difficult task and has not been achieved as of yet \cite{movileanu2009interrogating}. In fact, longitudinal ionic transport detects a current blockade which is the convolution of several blockade events from different amino acids \cite{branton2008potential}.

Transverse electronic transport, a technique in which amino acids are detected by a pair of electrodes transverse to peptide translocation, has been shown to be successful in identifying single amino acids and even in differentiating between tyrosine and phosphotyrosine \cite{ohshiro2014detection}, a post-translational modification. However, only 12 of the 20 amino acids were able to be detected by this technique with two different electrode gap distances (0.55 nm and 0.7 nm) since the tunneling current is highly dependent on this gap distance and an amino acid's ability to enter the gap. In other words, a single gap cannot be used for all amino acids.

This brings us to our proposed technique, sequencing proteins with {\it transverse ionic transport}. Like the two aforementioned techniques, this method is inspired by a DNA sequencing method \cite{wilson2013single,menard2012device} and does not require reagents or fractionation since these devices do not place a limit on the length of the polypeptide \cite{branton2008potential}, meaning these nanopore techniques have the potential to be much faster. The structure of this proposed device is the same as in \cite{wilson2013single,menard2012device}, with a longitudinal nanochannel for polypeptide translocation and an intersecting transverse nanochannel for ionic transport driven by an electric field, $E_{\perp}$, as in Fig. \ref{fig:schematic}. However, the longitudinal nanochannel must be larger than in \cite{wilson2013single} to accomodate the various sizes of the amino acids and instead of 4 DNA bases we need to distinguish 20 amino acids. Therefore, the molecular dynamics (MD) simulation method utilized in \cite{wilson2013single} is time prohibitive for our purposes so we resort to a hard sphere model to account for the electrostatic properties of each amino acid, which requires only one MD run per amino acid to execute. Afterwards we use Monte Carlo sampling to calculate ionic current distributions based on external azimuthal rotations ($\phi '$) and dihedral angle ($\phi$ and $\psi$) distributions, or Ramachandran plots. We show that the distribution of ionic currents for each of the 20 proteinogenic amino acids encoded by eukaryotic genes is indeed statistically distinct, and propose a protocol for {\it de novo} protein sequencing based on this technique.
\section{Theoretical Approach}
\label{sec:approach}
Let us then consider the configuration of crossed nanochannels we have in mind. Although not necessary for our conclusions, we assume for simplicity the nanochannels to have circular cross sections. We will discuss the suggested experimental preparation later in the manuscript, in the section titled Sequencing Protocol.

The polypeptide of interest unfolds inside a nanochannel pulled with a longitudinal force, while it blocks the ionic current flowing in a transverse channel, as schematically shown in Fig. \ref{fig:schematic}.

It is well understood that the hydration layers surrounding each amino acid have different binding energies \cite{wolfenden1981affinities,chang2007solvation}, which certainly affect the ionic transport transverse to each amino acid. In addition, the amino acid may attract or repel ions due to its solvated charge or polarity state \cite{kish2003na+,rulisek2000theoretical}. In order to understand the aqueous environment of each amino acid and determine its effect on the ionic transport, we run MD simulations for each amino acid. We consider the system at normal human body temperature, 310 K, and the solvated system is large enough to make quantum effects negligible. This allows us to use classical molecular dynamics and employ the highly-parallel NAMD2 \cite{Phillips2005} to run all of our simulations.

The MD setup starts with a single amino acid isolated from a straight (dihedral angles $\psi = \phi = 180^{\circ}$) peptide chain, as in Fig. \ref{fig:schematic} with proline (PRO) as an exception, which is positioned so that the $z$-axis is the longitudinal axis. The rest of the MD methods can be found in the Supporting Information.

The water padding is large enough in this system to examine proximal radial distribution functions (pRDFs) from the amino acid's surface for ${\rm K}^+$ and ${\rm Cl}^-$ up to the point where the concentrations level out to the bulk values. We use the radius from the surface of the amino acid because the features in the concentration will be more prominent as opposed to using the radius from the origin, since the amino acids have irregular shapes. Similarly calculated pRDFs on DNA have been shown to be fairly accurate for reconstructing the surrounding solute even when combining all surface atoms' pRDFs into one \cite{rudnicki1997modeling,feig1999sodium}, as is done in our calculations. 

To obtain the pRDFs, we count the number of ions (for ${\rm K}^+$ and ${\rm Cl}^-$) in 0.5 {\AA} thick shells starting from the surface of each amino acid, which is defined by the intersection of the composing atoms' van der Waals (vdW) spheres. We then calculate the volume of each shell by subtracting the inner volume of the intersecting spheres from the outer volume, using a grid approximation with 0.1 {\AA} sides for each volume calculation. With the number of ions and the volume of the corresponding shell we calculate the local concentration of ${\rm K}^+$ and ${\rm Cl}^-$ as a function of $r_{>}$, taken to be the perpendicular distance from the vdW surface to the radial midpoint of the shell, from the first shell at $r_{>} = 0.25$ {\AA} to the last at $r_{>} = 44.75$ {\AA}, which is below the 4.8 nm upper bound of water padding.
\begin{figure}[t]
\centerline{\includegraphics[width=.4\textwidth]{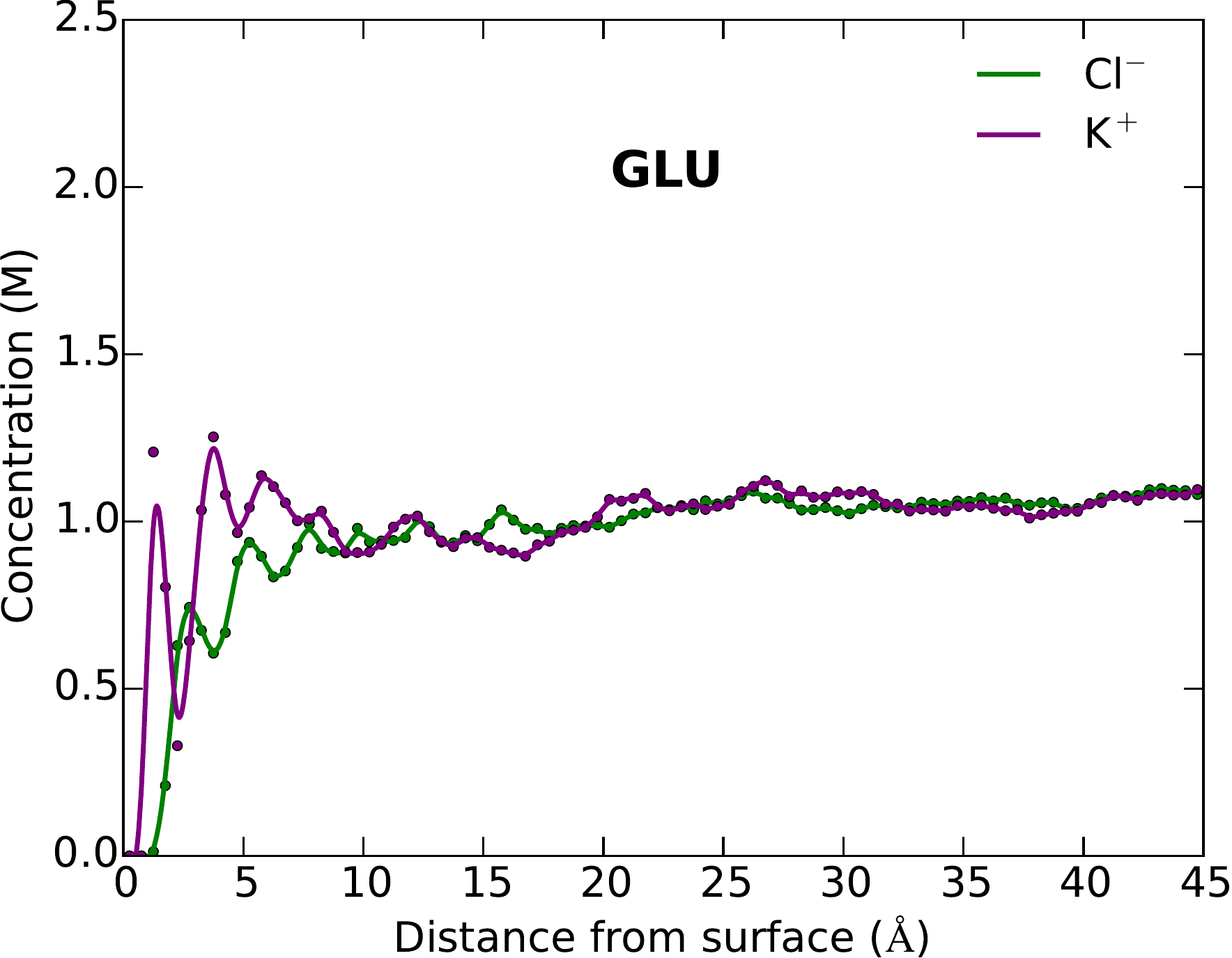}}
\centerline{\includegraphics[width=.4\textwidth]{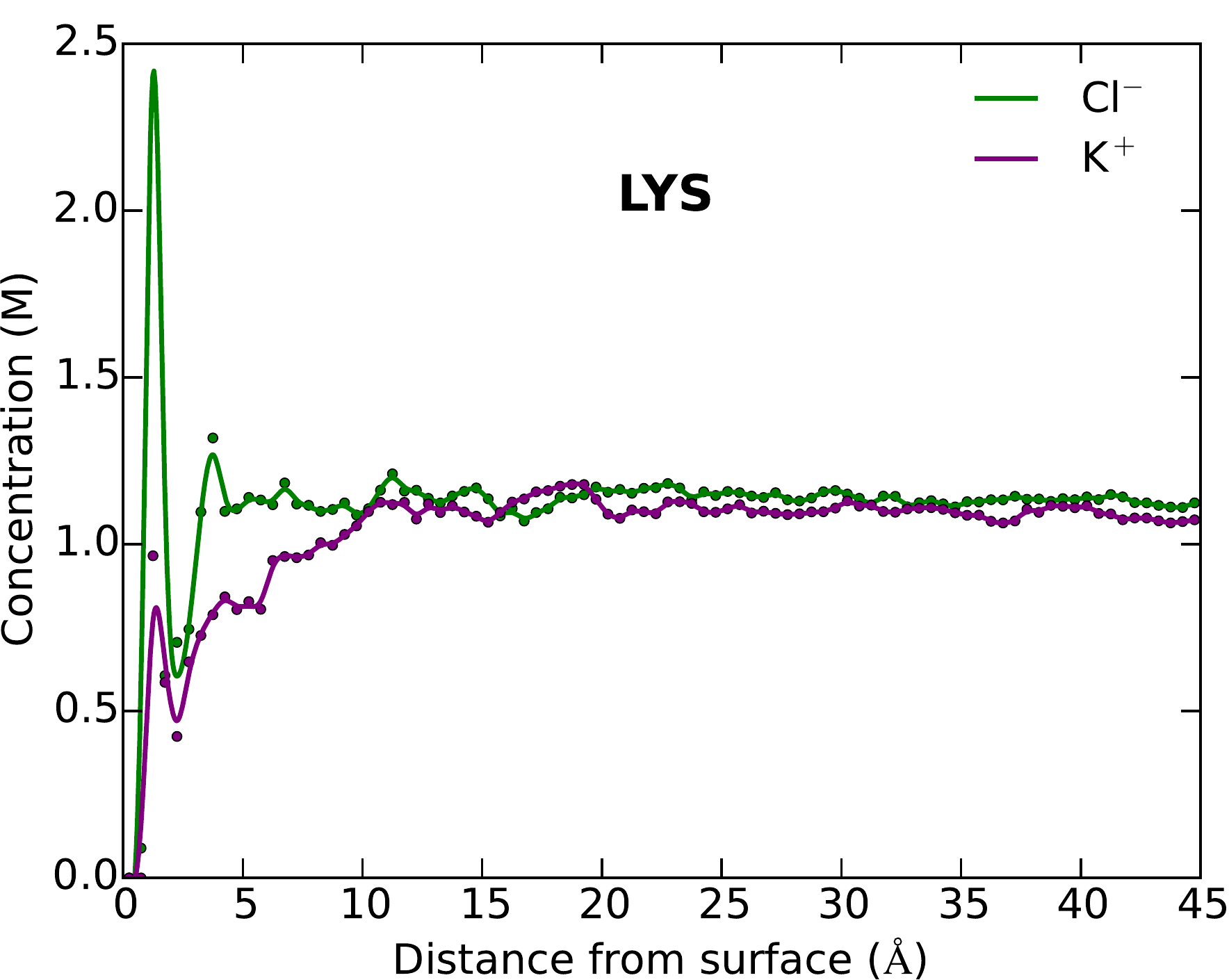}}
\centerline{\includegraphics[width=.4\textwidth]{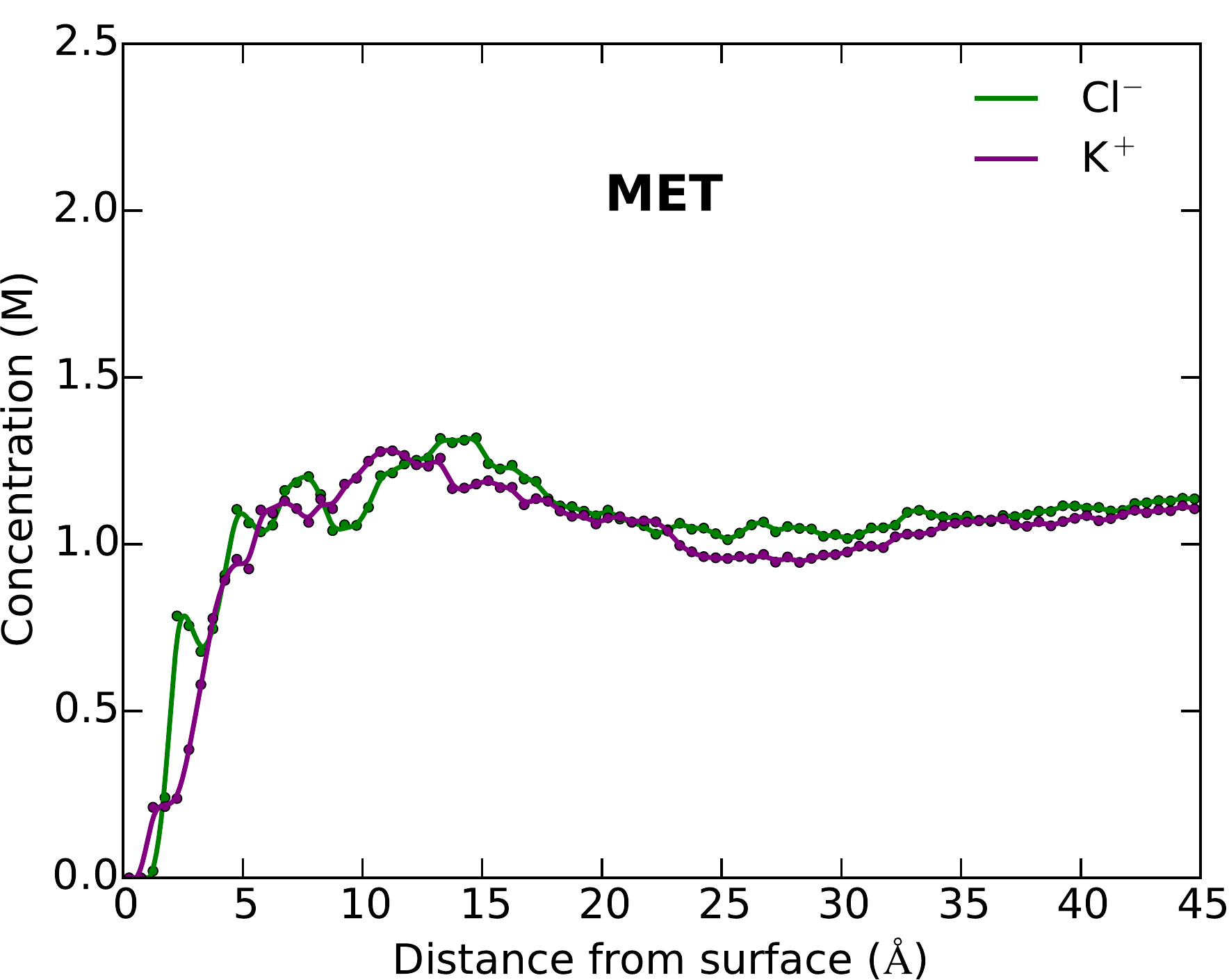}}
\caption{\label{fig:conc}(Color online) Plots of ionic concentration against distance from each amino acid's vdW surface, $r_{>}$, for amino acids GLU, LYS, and MET. ${\rm K}^+$ is represented by the purple line and ${\rm Cl}^-$ is represented by the green line.}
\end{figure}

As can be seen in Fig. \ref{fig:conc}, the concentrations reach a sufficiently steady bulk value at varying radii, with the maximum bulk $r_{>}$ determined to be approximately 15 {\AA}. Therefore we can focus on the part of the plots pertaining to $r_{>} \leq 15$ {\AA} to determine the solvation properties of each amino acid. As an example of our numerical procedure, we have chosen to feature the amino acid GLU in Fig. \ref{fig:conc}A, which has a negatively charged side chain at physiological pH (7.4), lysine (LYS) in Fig. \ref{fig:conc}B, which has a positively charged side chain at the same pH, and methionine (MET) in Fig. \ref{fig:conc}C, whose side chain is hydrophobic at this pH. These three amino acids are of similar size, which allows us to better compare the effects of charge states on transverse ionic current. We can immediately notice that the part of the pRDFs that we care about is quite different for each featured amino acid. GLU in Fig. \ref{fig:conc}A has a higher concentration of ${\rm K}^+$ due to its negativity while LYS in Fig. \ref{fig:conc}B has a higher concentration of ${\rm Cl}^-$ due to its positivity. Then there is the hydrophobic MET in Fig. \ref{fig:conc}C, which appropriately repels both ${\rm K}^+$ and ${\rm Cl}^-$ without much preference.

In the setting of an external electric field driving transverse ionic flow around an amino acid within a peptide, the potential barrier that ions must overcome in transport is influenced mostly by the electric potential in the neighborhood of the area-limiting cross section perpendicular to the ionic flow, imaged in Fig. \ref{fig:schematic} as the black thick-dashed line. This is partly due to the short interference time between the flowing ions and the circumvented amino acid. In our theoretical approach, we treat the equilibrium ionic concentrations as indicators of this electric potential to develop a hard sphere model with which we can calculate the distribution of ionic current for each amino acid. By calculating an effective radius, $r_{\rm eff}$, that is applied to every atom in the amino acid beyond its vdW radius, we can sample many amino acid orientations using a Monte Carlo approach to determine all of the ionic current distributions. We theorize that most of the variation in the transverse ionic transport will come from the exclusionary effects of the amino acid with respect to the direction of ionic flow, meaning that a large pool of orientations must be sampled to obtain an accurate view of these distributions.

In order to obtain the effective radius for each amino acid, we start with the definition of the average transverse ionic current of an ionic species $i$, ${\rm K}^+$ or ${\rm Cl}^-$ in our case, with valency $\tilde{z}_i$ flowing around an amino acid.
\begin{equation}
\begin{split}
\langle I_i \rangle &= q \tilde{z}_i \int^{r_f}_{r_i}{\int^{\theta_f}_{\theta_i}{\langle \tilde{g}_i \tilde{v}_i(r, \theta '(\hat{r}'), \phi '(\hat{r}')) \rangle_{\hat{r}'} r {\rm d} \theta}{\rm d} r} \\
&= C \int^{r_f}_{r_i}{\langle \tilde{g}_i \tilde{v}_i \rangle_{\hat{r}'}(r) r {\rm d} r} \\
&= C' \int^{r_f}_{r_i}{\langle \tilde{g}_i \tilde{v}_i \rangle_{\hat{r}'}(r) \frac{\tilde{A}(r)}{2r} {\rm d} r} \\
&\approx C'' \int^{r_{\rm b}}_{0}{g_i(r_{>}) v_i(r_{>}) \frac{A(r_{>})}{2(r_{>} + r_{\circ})} {\rm d} r_{>}} \\
&= C'' \int^{r_{\rm b}}_{r_{\rm eff}}{g_{i , {\rm b}} v_{i , {\rm b}} \frac{A(r_{>})}{2(r_{>} + r_{\circ})} {\rm d} r_{>}}
\end{split}
\label{eq:rad_approx}
\end{equation} 
Here, the identifying transverse ionic current, $I_i$, through the aforementioned area-limiting cross section perpendicular to the ionic flow, is averaged over all rotational orientations equally with $\hat{r}'$ representing the unit $\vec{r'}$ vector of the amino acid while $q$ is the electron charge and $C$, $C'$, and $C''$ are constants. $[\theta_i , \theta_f]$ is the window of $\theta$ where the amino acid under study has non-negligible influence compared to neighboring amino acids. $r_i$ is the radius where $\tilde{g}_i$, the local number density as a function of spherical coordinates, is first nonzero. $r_f$ is the radius where the influence of the amino acid is no longer felt in the concentration and thus we need not continue the integral for the purpose of the effective radius, $r_{\rm eff}$, calculation. $\tilde{v}_i$ is the transverse velocity through the cross section as a function of standard spherical coordinates while $\tilde{A}(r)$ is the surface area of the sphere of radius $r$. In addition, $r_{>}$ is the perpendicular distance from the vdW surface to the radial midpoint of a shell of thickness ${\rm d}r_{>}$ and surface area $A$, $r_{\circ}$ is the average radius from the origin to the vdW surface, $r_{\rm b}$ is a value of $r_{>}$ where the pRDF, $g_{i}$ (plotted in Fig. \ref{fig:conc}), has become sufficiently steady around the bulk density, $g_{i , {\rm b}}$, so as to represent a shell in the bulk, $v_i$ is the flow velocity of the ion species $i$ as a function of $r_{>}$, while $v_{i , {\rm b}}$ is the maximum of $v_i$, which occurs in the bulk by construction.

The first approximation that we make is that all of the rotational orientations are uniformly likely, when in reality $\theta '$ is fairly constant due to the stiffness of the peptide bond and given how small the diameter of the pore is in comparison to the length. However, when we average over $\phi '$ we fully explore the number density around the shell, so averaging over $\theta '$ does not introduce any new data but adds more weight to the side chain as opposed to the ends of the backbone. This counteracts the simplification we make in our MD runs where we use isolated amino acids and include the number density at the ends of the backbone, which would normally be expelled by the nearest neighbor amino acids. Also, the internal dihedrals are assumed fixed since they do not fluctuate much under the imposed longitudinal electric field (see their implementation in the current distribution calculations). Lastly, when we change variables from $r$ to $r_{>}$ we have to approximate $r$ as $r_{>} + r_{\circ}$, which is a minor approximation when considering that all of the other functions in the integral have well-defined transformations. We can now use the following simplified equation to calculate the effective radius for our hard sphere model for every amino acid and ion species combination:
\begin{equation}
\begin{split}
&\int^{r_{\rm b}}_{r_{\rm eff}}{\frac{A(r_{>})}{2(r_{>} + r_{\circ})} {\rm d} r_{>}} = \\
&\qquad \qquad \qquad \int^{r_{\rm b}}_{0}{\frac{g_i(r_{>})}{g_{i , {\rm b}}} \frac{v_i(r_{>})}{v_{i , {\rm b}}} \frac{A(r_{>})}{2(r_{>} + r_{\circ})} {\rm d} r_{>}} .
\end{split}
\label{eq:eff_radius}
\end{equation}

However, this equation requires the ratio of the transverse flow velocity compared to the bulk, and due to the small length scales we can use the Stokes equation, similar to \cite{qiao2003ion}. The details of this calculation can be found in the Supporting Information. From these calculations we find that $r_{\rm b} = (R - r_{\circ}) / 2$ and then from our pRDF plots (see Fig. \ref{fig:conc}) we learn that the bulk concentrations start at approximately $r_{\rm b} \geq 15$ {\AA}. Therefore for our model to work we have to take $R \geq 30 + \max \{ r_{\circ} \} = 34.16$ {\AA}, where the max is over all amino acids, and then in the interest of minimizing the bulk ionic current we choose $R = 35$ {\AA}. We also set the transverse nanochannel radius to the same value for simplicity.
\begin{figure}[t]
\centerline{\includegraphics[width=.4\textwidth]{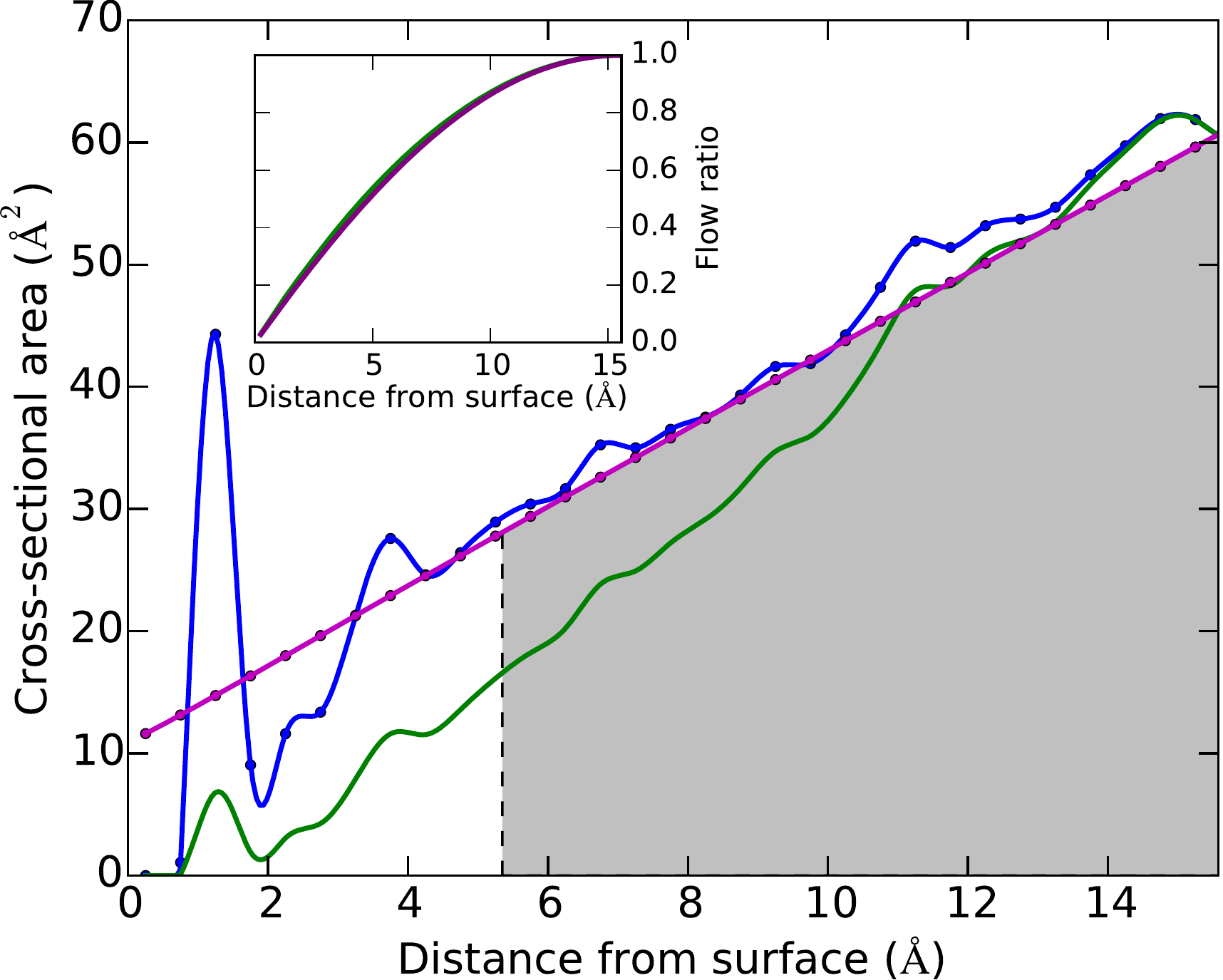}}
\centerline{\includegraphics[width=.4\textwidth]{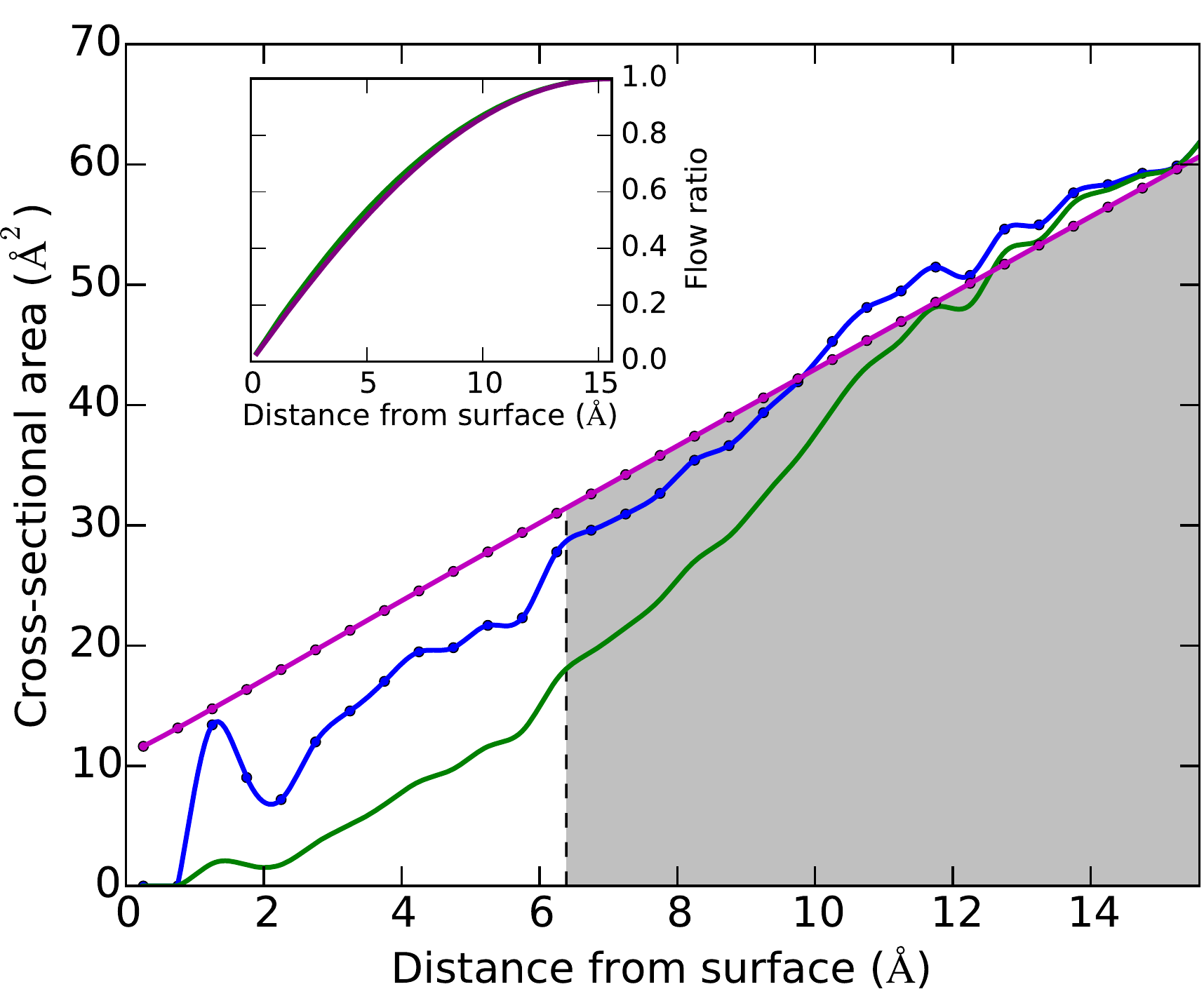}}
\caption{\label{fig:area}(Color online) The top graph represents area plots of ${\rm Cl}^-$ around LYS while the bottom graph shows area plots of ${\rm K}^+$ around LYS. The straight magenta line is the average cross-sectional area that the shell of thickness $\Delta r_{>} = 0.5$ {\AA} at $r_{>}$, the distance from the vdW surface of LYS, occupies in the plane $y = 0$. The blue line represents the average cross-sectional area that the ionic solution, with the number of ions from the shell of thickness $\Delta r_{>} = 0.5$ {\AA} at $r_{>}$, would occupy in the plane $y = 0$ if those ions were reorganized to have bulk concentration, $g_{i , {\rm b}}$. The smooth green curve is the blue curve modulated by the ratio of the velocity with its maximum, $v_i / v_{i , {\rm b}}$, which is plotted in the inset of each graph. The area under the smooth green curve is equal to the shaded gray area under the straight magenta curve while the dashed vertical line marks the effective radius for ion species $i$ specifically for LYS.}
\end{figure}

The insets of Fig. \ref{fig:area} show the results of our calculations for $v_i / v_{i , {\rm b}}$; the top graph represents ${\rm Cl}^-$ around LYS while the bottom graph shows ${\rm K}^+$ around LYS. The other amino acids have similar parabolic forms for $v_i / v_{i , {\rm b}}$, but differing $r_{\rm b}$ because of differing $r_{\circ}$. With $v_i / v_{i , {\rm b}}$ calculated for every amino acid we can return to Eq. \eqref{eq:eff_radius} to calculate our effective radii for our hard sphere model. This calculation is shown graphically in Fig. \ref{fig:area}, where the straight magenta line is the argument (including ${\rm d} r_{>}$ as $\Delta r_{>} = 0.5$ {\AA}) of the left-hand side of Eq. \eqref{eq:eff_radius}, which is the average cross-sectional area that the shell of thickness $\Delta r_{>}$ at $r_{>}$ occupies in the plane of interest ($y = 0$). The blue line represents the argument of the right hand side of Eq. \eqref{eq:eff_radius}, again including ${\rm d} r_{>}$ as $\Delta r_{>} = 0.5$ {\AA} without the modulation of the velocity ratio, leaving the average cross-sectional area that the ionic solution, with the number of ions from the shell of thickness $\Delta r_{>}$ at $r_{>}$, would occupy in the plane of interest ($y = 0$) if those ions were reorganized to have concentration $g_{i , {\rm b}}$. Finally the smooth green curve is the blue curve modulated by $v_i / v_{i , {\rm b}}$. The area under the smooth green curve is equal to the shaded gray area under the straight magenta curve, with the dashed vertical line marking not only where the shaded gray area ends on the left but also the effective radius for ion species $i$ for the given amino acid. Because of the influence of the velocity, the fluctuations in concentration farther from the amino acid have more effect than closely bound spikes. For example, the ${\rm Cl}^-$ ion atmosphere located 1 {\AA} from the surface of LYS has less effect on the effective radius compared to the next spike in concentration further out from the amino acid, as seen in the green curve. The fact that LYS is positively charged still shows in the effective radii though, with the attractive ${\rm Cl}^-$ ions having a 5.38 {\AA} addition to the vdW surface compared to 6.43 {\AA} for the repulsive ${\rm K}^+$ ions. The rest of the effective radii can be found in Table \ref{table:eff_rad} of the Supporting Information.

This brings us to our Monte Carlo calculation of the transverse ionic current around each amino acid. Now that we have $r_{\rm eff}$ for each ion species that we add to the vdW radius of every atom in our amino acid, we can compare the available cross-sectional area through the $y = 0$ plane and apply the same bulk concentration and estimated bulk velocity, $g_{\rm b} = 1$ M and $v_{\rm b}$, to all amino acids to obtain the ionic current values. We do not need to evaluate the available area in the entire cross section though since we only need to calculate up to the largest radius determined by $r_{\rm eff}$ for all amino acids. Therefore we use a radius of $R / 2$ from the origin (see Fig. \ref{fig:schematic} where we are now limited to $r = R$) as the circular boundary for all of the amino acids since this circle encloses all of the extended amino acid surfaces in any applicable rotational configuration while also being enclosed by the bulk boundary defined by $r_{>} = r_{\rm b}$ where the velocity begins to decline from $v_{\rm b}$. We also approximate $[\theta_i , \theta_f]$ as $[ \pi / 4 , 3 \pi / 4 ]$ by comparing the backbone ends' vdW radius to half of the distance in $z$ between amino acids (half of ideally $\sim 3.8$ {\AA} \cite{stirnemann2014force}). In this manner we can ignore portions of the cross section that would clearly be dominated by neighboring amino acids for the purposes of understanding each amino acid's transverse ionic transport signature.

As previously mentioned, the current becomes sensitive to rotational conformations and dihedral angles in this portion of the calculation. Therefore, instead of assuming uniformity in $\theta '$ and straight dihedral angles like we did for the effective radius, we fix $\theta '$ to 0 due to the rigidity of the peptide bond and we use Ramachandran plots, \cite{stirnemann2013elasticity,ramachandran1968conformation}, to sample realistic values for $\phi$ and $\psi$, dihedral angles, which encompass the internal degrees of freedom for a chain of amino acids \cite{ramachandran1968conformation,richardson1981anatomy}. That leaves the azimuthal angle, $\phi '$, which we leave as uniformly distributed since as a whole the peptide does not have an azimuthal preference, except if the peptide is very short in which case the transverse electric field that is only applied to a few amino acids can affect the entire chain. We then apply Monte Carlo to a lone amino acid, the details of which can be found in the Supporting Information. The reason we use a lone amino acid, the same one from our MD simulations, for calculating the ionic current distributions is that the first step to understanding the viability of this technique is distinguishing each amino acid separately via transverse ionic current. Since most of the exclusion due to the amino acid comes from the region of small $z$, where the uniqueness of the amino acid is demonstrated, the exclusion from one amino acid in a chain can be derived from our single amino acid distributions. As a result we do not treat the effect of neighboring PRO, which alters the dihedral angles so as to straighten the polypeptide chain. However, changing an amino acid's dihedrals slightly does not change the ionic current distributions much since most of the variation in the current comes from azimuthal rotation of the amino acid.

Lastly, we must calculate the bulk velocity, $v_{\rm b}$, that we will use in the simple equation for the transverse ionic current, $I_i = q \tilde{z}_i g_{\rm b} v_{\rm b} \langle A_i \rangle$ and $I = \sum_i I_i$, where $\langle A_i \rangle$ is the average area outside of the effective surface from Monte Carlo. This calculation can be found in the Supporting Information, resulting in $v_{\rm b} = 77.23$ m/s.
\begin{figure*}[t!h]
\begin{center}
\centerline{\includegraphics[width=.7\textwidth]{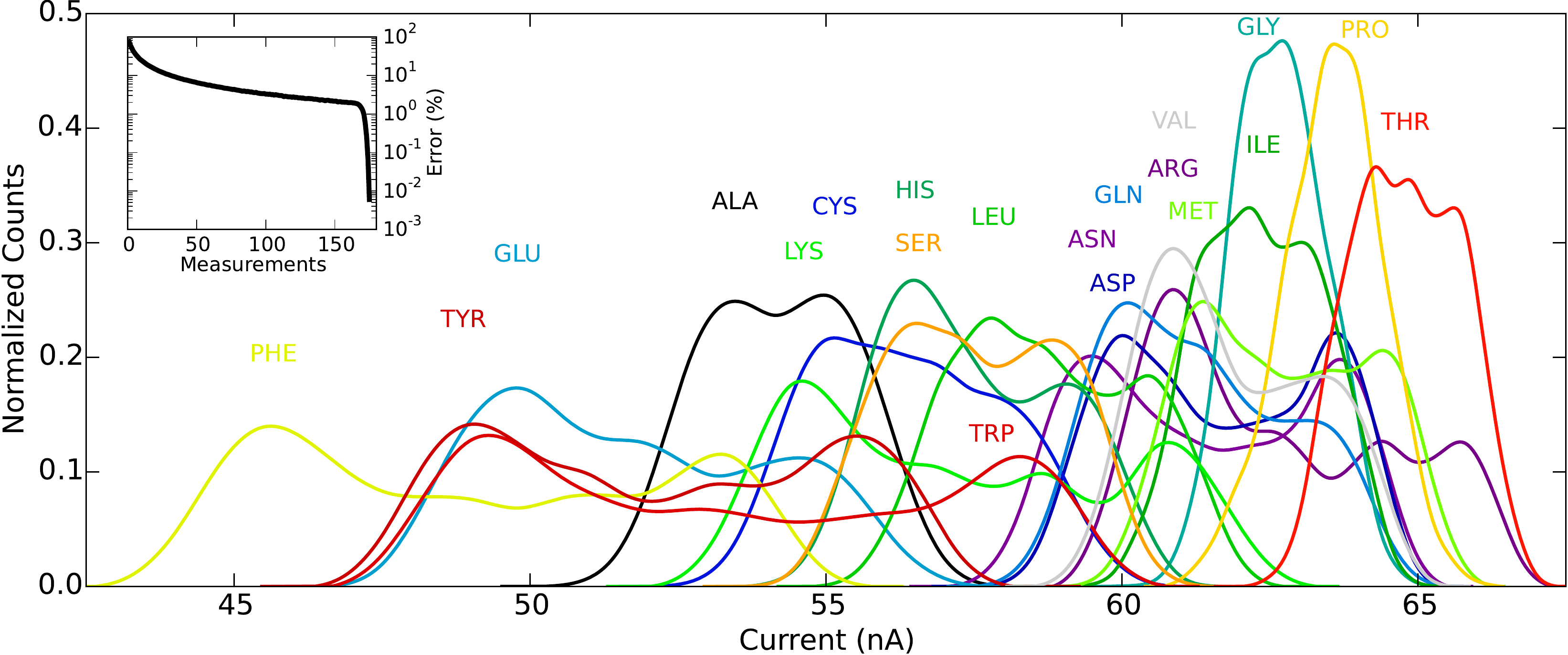}}
\caption{\label{fig:distr}(Color online) The transverse ionic current distributions for all 20 proteinogenic amino acids encoded by eukaryotic genes (identified with their standard three-letter abbreviations). The distributions have been normalized to the current values in nA. The inset plots the average error percentage over all 20 amino acids of identifying an amino acid correctly using $M$ current measurements from that amino acid where the error percentage is on a log scale.}
\end{center}
\end{figure*}
\section{Results and Discussion}
\label{sec:results}
With a set of ionic currents for each amino acid determined from Monte Carlo utilizing our hard sphere model, we histogram each set of currents and use cubic spline interpolation to arrive at Fig. \ref{fig:distr}. The ionic currents tend to form multimodal (most often bimodal) distributions that are best described as a mixture of several normal distributions. The first and last peaks of each distribution tend to be the highest due to the variation in $\phi '$. This is because the ionic current as a function of $\phi '$ is roughly sinusoidal with a period of $\pi$ and $\phi '$ is uniformly distributed, which means the near minimum and near maximum values of the ionic current are chosen the most. Also due to the size of the nanochannels, the ionic current ranges in the tens of nA, which is well within the range of modern measurement devices that can resolve pA currents \cite{gao2014integrated,menard2012device}. Beyond that, this ionic current only represents up to $R / 2$ of the whole cross section. By using the parabolic $\hat{v}$ from the bulk region (see the end of Sequencing Protocol) we calculate the contribution from the rest of the cross section, $r_{>} > r_{\rm b}$ but still within the $\theta$ limitations, as 69.86 nA after correcting the velocity for experiment. This value is comparable to the ionic current values from Fig. \ref{fig:distr}, meaning the distinctive component of the ionic current will not be dwarfed by the bulk in an experimental setting.

Although a fair number of amino acids do not deviate much from their vdW size identity, namely PRO remaining on the smaller side (large current) and phenylalanine on the larger side (small current), many more ({\it e.g.}, alanine) have shifted due to their interaction with the ions. However, the vdW volume does remain strongly relevant in the standard deviation of the distributions, where the larger amino acids (arginine, phenylalanine, tryptophan, tyrosine) find more variation in ionic current as the dihedrals or $\phi '$ are altered.

At a glance there is significant overlap between all of the distributions, yet the graph seems crowded mostly because of the sheer amount of plots to compare. We quantify the distinguishability of the ionic current distributions by calculating the error in selecting the correct amino acid, $X$, given $M$ measurements from $X$. Based on the maximum likelihood decision rule \cite{duda2012pattern}, the error is defined by
\begin{equation}
\begin{split}
&e^{X}_{m} = 1 - \frac{1}{J}\sum^{J}_{j = 1} \left\lfloor \frac{1}{19} \sum^{\{Y\}}_{Y \neq X} H \left( \prod^{M}_{m = 1} P^{X}(I^{X}_{m,j}) - \right. \right.\\
&\hspace{12.5em} \left. \vphantom{\sum^{\{Y\}}_{Y \neq X}} \left. \prod^{M}_{m = 1} P^{Y}(I^{X}_{m,j}) \right) \right\rfloor ,
\end{split}
\label{eq:error}
\end{equation}
where $J$ is the total number of realizations of the error calculation, $\{Y\}$ is the set of all 20 amino acids, $H$ is the Heaviside step function, and $P^{Y}(I^{X}_{m,j})$ is the probability of $I^{X}_{m,j}$, the $j$th realization of the $m$th ionic current measurement sampled from the current distribution for $X$, in $Y$'s ionic current distribution. Here, we assume that each measurement of ionic current is approximately independent. Next we average over $X$ to obtain $\langle e^{X}_{m} \rangle_{X}$ and then multiply by 100 to get the error percentage, which is plotted in the inset of Fig. \ref{fig:distr}. The error drops at a moderate rate with increasing $M$, but significantly drops off for $M > 160$ when the likelihood of at least one measurement giving zero probability to incorrect amino acids becomes very likely, making the product of those incorrect probabilities zero. For instance, at $M = 175$ the error percentage is practically 0\%, and certainly less than 0.1\%, a reasonable level of error. With a measurement frequency of 100 kHz, \cite{gao2014integrated}, and a best case scenario of 175 measurements per residue without any lapses in between, the sequencing rate becomes 571 residues per second. 
\section{Sequencing Protocol}
\label{sec:experiment}
To build a nanofluidic device with intersecting channels as we suggest one may employ focused ion beam milling, as achieved in \cite{menard2012device} with two 10 nm diameter intersecting nanochannels. Our model requires two 7 nm diameter intersecting nanochannels, which is certainly achievable given that \cite{menard2010fabrication} has shown non-intersecting sub-5 nm nanochannels from the focused ion beam milling technique. Although we have predicted that all 20 amino acids are statistically distinct within the framework of circular channels, other cross sections like rectangles or ellipses for the transverse channel allow fewer amino acids to blockade the ionic transport but still provide enough space for ions to flow past the translocating polypeptide. This results in improved residue selectivity and therefore decreased error as well as reduced post-processing time for deconvolution of the amino acid signals, which is necessary if more than one amino acid resides in the nanochannel intersection. Since the source of the distinguishability of the amino acids is their structural and electronic uniqueness we can assume that using a rectangular or elliptical transverse cross section with enough space along $x$ for ionic flow would also result in 20 statistically distinct amino acids.

Once the sequencing device is built with transverse electrodes to control ionic flow, the protein or polypeptide of choice must be unfolded to translocate it through the longitudinal nanochannel. By using a high enough pulling force, around 250 pN \cite{carrion1999mechanical,stirnemann2014force} that we also apply to our model, the polypeptide will unfold as well as translocate through the nanochannel. As opposed to chemical denaturing, force unfolding results in more confined and reliable Ramachandran plots \cite{carrion1999mechanical,stirnemann2014force}, which directly translates to more reliable ionic current distributions. After the polypeptide is unfolded the pulling force can be adjusted according to one's ionic current measurement frequency and desired rate of error. For example, a desired 0.1\% or less of error requires $M = 175$ and with a sequencing rate of 100 kHz as before, the maximum pulling speed would be 217 nm/s assuming an amino acid length of 3.8 {\AA}. As a result, the maximum applicable pulling force would be $\sim 180$ pN \cite{carrion1999mechanical}.

The next issue is then how this polypeptide is pulled through the nanochannel. As we have discussed, amino acids have varied charge states in solution. Therefore, to utilize an electric field for pulling (see Fig. \ref{fig:schematic}) one has to attach charges to the polypeptide. These charges must be attached at the end of the chain so that one does not interfere with the ionic transport signatures of each amino acid. The best way to achieve this is by using a combination of solid phase peptide synthesis (SPPS), which excels at synthesizing smaller peptides \cite{merrifield1963solid}, and native chemical ligation (NCL) \cite{dawson1994synthesis} to attach a sequence of charged amino acids to the N-terminus of the polypeptide under study. We choose GLU as our charged amino acid because of how easily differentiable it is from the other amino acids (see Fig. \ref{fig:distr}) and how easy it is to produce. Using Fmoc, 9-fluorenylmethyloxycarbonyl or the chemical group that protects the N-terminus from reactions until desired, SPPS starting with N,N-bis(2-mercaptoethyl)-amide (BMEA) \cite{hou2010peptidyl} one creates a sequence of GLU with a length that will give the polypeptide chain plus GLU sequence a large enough charge to pull with an electric field. Fmoc SPPS is also used to attach a CYS residue to the N-terminus of the unknown polypeptide with a polyethylene glycol (PEG) support \cite{roberts2012chemistry}. Then one uses NCL to take advantage of the transthioesterification reaction to form a native amide bond between the N-terminal CYS residue and the thioester precursor BMEA \cite{hou2010peptidyl}.

Another option is to use optical tweezers \cite{ashkin1986observation,grier2003revolution} to target a terminal amino acid to pull the whole polypeptide. This approach has been utilized for longitudinal nanopore DNA sequencing \cite{keyser2006direct,trepagnier2007controlling}, resulting in more control over translocation due to the high tunability of optical tweezers. Advances in optical tweezers further allow a single beam to trap multiple targets \cite{arai2004synchronized}, potentially with computer-generated holograms \cite{grier2006holographic}, which would allow even more control over the entire polypeptide.
\section{Summary}
\label{sec:summary}
We have proposed a novel {\it de novo} protein sequencing method in which an unfolded protein confined to a nanochannel is probed by transverse ionic transport through an intersecting nanochannel. This method promises to offer improved discrimination between amino acids by utilizing the 3-dimensional structure and electronic properties of each amino acid, as compared to techniques like mass spectrometry that can only probe total mass and charge \cite{standing2003peptide}. We developed a hard sphere model for transverse ionic transport that employs the average equilibrium ionic concentrations surrounding all 20 amino acids derived from MD and ionic flow ratios determined by the Stokes equation. With this hard sphere model we were able to calculate distributions of ionic current for each amino acid based on Monte Carlo sampling of internal and external rotational conformations. All 20 amino acids were found to be statistically distinct and a sequencing error rate per residue of less than 0.1\% was obtained with $M = 175$ measurements per amino acid, implying a best case scenario of 571 residues per second with a measurement frequency of 100 kHz \cite{gao2014integrated}.

This approach is certainly experimentally achievable since 10 nm diameter intersecting nanochannels have been demonstrated for the purpose of DNA sequencing \cite{menard2012device} and polypeptides can be pulled through the nanochannel with optical tweezers or by adding charged residues to the polypeptide terminus and employing an electric field. Protein sequencing is very important since DNA sequencing cannot predict post-translational modifications and the ability to identify the sequence of a protein leads to the ability to understand its structure, which is the key to understanding many crippling diseases like Alzheimer's \cite{kelley2008simulating}. We therefore hope our work will motivate the experimental realization of the proposed protein sequencing protocol.
\bibliography{all_bibs}
\clearpage
\section{Supporting Information}
\label{sec:si}
\renewcommand\theequation{S\arabic{equation}}
\setcounter{equation}{0}
\paragraph{Molecular Dynamics}
The amino acid is centered along the $z$-axis according to the geometric center in $z$ of its terminal N and C atoms, while the molecule is centered in the $xy$-plane according to the geometric center in $x$ and $y$ of its terminal N atom and a nearest neighboring amino acid's terminal N atom. To fix the rotation angle between amino acids, as a convention, the terminal N atoms always have $y = 0$, as is the case in Fig. \ref{fig:schematic}. 

Since PRO has more rigid dihedral angles, we need to center it with the help of two neighboring GLYs, which have flexible dihedral angles, on each side of a single PRO. The nearest neighbor GLYs are configured to have dihedral angles that compensate for those of PRO while the farthest neighbor GLYs are configured to be straight, so that PRO and the two straight GLYs are directed along the longitudinal axis while the two straight GLYs are aligned in the $xy$-plane. As a result, we can center and then isolate PRO by using the usual centering method on the geometric average of the two straight GLYs, staying consistent with the choice of angles for the rest of the amino acids.

Once the amino acid is isolated, we solvate the system into a right hexagonal prism with regular hexagonal $xy$-planes having a height of 11 nm and an apothem of 5.9 nm to be used in NAMD2 with periodic boundary conditions in all three dimensions of space. This configuration gives every atom from every amino acid at least 4.8 nm of water padding in the unit cell, or in other words at least 9.6 nm of water between any atom and the closest atom in any neighboring periodic image. We then passivate and ionize the system to about 1 M of KCl, a typical biological solute. The size of the ions will certainly change the average local concentrations near the amino acid, which may then affect the ionic transport. We utilize the CHARMM22 with CMAP force field \cite{brooks2009charmm,mackerell2004extending} for all of the amino acid, TIP3P water, and ion interactions. Each amino acid was held fixed throughout the run so that it would not diffuse around and the surrounding solution could equilibrate and be analyzed consistently. After equilibrating at 0 K and progressively ramping up the temperature to 310 K, the system is allowed to evolve in an NPT ensemble first for 1 ns followed by an NVT ensemble for 5 ns, all with 1 fs time steps and 1 ps coordinate recordings. The temperature is held fixed using a Langevin thermostat with a damping coefficient of 5 ${\rm ps}^{-1}$. The first ns of the NVT production run is discarded as transient, leaving 4 ns of run time, or 4000 coordinate snapshots, to analyze radial concentration profiles.

\paragraph{Velocity calculation}
Due to the small length scales of the intersecting portion of our nanochannel system, we can use the Stokes equation,
\begin{equation}
\frac{\rm d}{{\rm d} x} \left ( \mu \frac{\rm d}{{\rm d} x} \hat{v}_i(x) \right ) + q \tilde{z}_i \hat{g}_i(x) E_{\perp} = 0 ,
\label{eq:vel_x}
\end{equation}
where $\mu$ is the dynamic viscosity of the fluid, $E_{\perp}$ is the external electric field applied in the $y$ direction, $\hat{v}_i$ is the transverse velocity through the cross section, $\hat{g}_i$ is the local number density, and ion-ion interactions are ignored. Here the flow velocity is independent of $y$ due to the fact that the transverse nanochannel length is much larger than the diameter of the longitudinal nanochannel, $2R$, and that $2R$ is comparable to the diameter of the transverse nanochannel (as depicted in Fig. \ref{fig:schematic}). Independence from $z$ is similarly due to the longitudinal nanochannel's length being much larger than $2R$ but we must also choose $R$ to be large enough for ions to diffuse along $z$ after they enter the longitudinal channel at $y = \pm R$. This will make any variation along $z$ have negligible impact on the end result of an average $v_i$ over all rotational conformations. In our case we simply use the dynamic viscosity of water ($\mu = 7.5 \times 10^{-4} \, {\rm Pa} \cdot {\rm s}$), even though the viscosity of water with ions will vary slightly \cite{qiao2003ion}, and a reasonable value of $E_{\perp} = 5 \times 10^8 \, {\rm N} / {\rm C}$ taken from \cite{qiao2003ion}. However, we must transform this equation into one that depends on $r_{>}$ to obtain $v_i$. Since $v_i$ and $g_i$ are averages over all rotational orientations, the problem is condensed to the region $x > 0$ and $[\theta_i , \theta_f]$. With $\theta_i$ and $\theta_f$ close enough to $\pi / 2$, we can approximate $r_{>}$ as $x - x_{\circ}$, where $x_{\circ}$ is some constant, since at least for the amino acid backbone the contour lines of $r_{>}$ resemble those of $x$. With these approximations we obtain
\begin{equation}
\frac{\rm d}{{\rm d} r_{>}} \left ( \mu \frac{\rm d}{{\rm d} r_{>}} v_i(r_{>}) \right ) + q \tilde{z}_i g_i(r_{>}) E_{\perp} = 0 .
\label{eq:vel_r}
\end{equation}
This will give us the rough form of $v_i / v_{i , {\rm b}}$ between the following boundary conditions:
\begin{equation}
\begin{split}
&v_i(r_{>} = 0) = 0 \\
&v_i(r_{>} = \bar{r}_{>} = \frac{R - r_{\circ}}{2}) \geq v_i(r_{>}) .
\end{split}
\label{eq:vel_bc}
\end{equation}
$\bar{r}_{>} = (R - r_{\circ}) / 2$ is approximately halfway between the vdW surface and the longitudinal nanochannel surface and is also our upper bound on $r_{>}$ as the domain of $v_i$. To obtain $v_i$ for the entire range necessary for Eq. \eqref{eq:eff_radius}, we require that $r_{\rm b} = \bar{r}_{>} = (R - r_{\circ}) / 2$ since $\bar{r}_{>}$ must be in the bulk as well. From our pRDF plots (see Fig. \ref{fig:conc}) we learn that the bulk concentrations start at approximately $r_{>} = 15$ {\AA}, meaning that $r_{\rm b} \geq 15$ {\AA}. Therefore for our model to work we have to take $R \geq 30 + \max \{ r_{\circ} \} = 34.16$ {\AA}, where the max is over all amino acids, and then in the interest of minimizing the bulk ionic current we choose $R = 35$ {\AA}.

\paragraph{Monte Carlo}
The Ramachandran plots that we use in our Monte Carlo calculations account for a 250 pN longitudinal force that is applied to the polypeptide chain (ubiquitin and polyglycine in \cite{stirnemann2013elasticity}) to pull it through the nanochannel. The pulling force acts to limit the phase space available to the dihedral angle pair $(\phi , \psi)$, making the configurations that are close to straight ($\psi = \phi = 180^{\circ}$) much more appealing \cite{stirnemann2013elasticity}. PRO and GLY have significantly different plots from the rest of the amino acids due to how the side chain of PRO bonds with its own amine nitrogen, part of the amino acid backbone, leading to restricted dihedrals while GLY has a hydrogen instead of a side chain leading to more freedom in the dihedrals. This way, while the rest of the amino acids are described by the Ramachandran plot of ubiquitin, which contains all 20 of the proteinogenic amino acids encoded by eukaryotic genes and well represents 18 of them, we describe PRO with the $(\phi , \psi)$ plot from the isolated PRO values within ubiquitin and GLY with the Ramachandran plot of the polyglycine analog of ubiquitin \cite{stirnemann2013elasticity}. With these Ramachandran plots we use Monte Carlo sampling to obtain $(\phi , \psi)$ pairs that we then implement on a lone amino acid, where the number of realizations is dependent on the size of the domain of the Ramachandran plot (1408 realizations for ubiquitin). We also rotate the amino acid in $\phi ' \in [ 0 , 2 \pi )$ by all multiples of $\pi / 12$. Then the amino acid is projected onto the $y = 0$ plane and using Monte Carlo (1000 realizations) we calculate the area outside of the effective surface, called $A_i$, yet within either $\pi / 2$ sector of radius $R / 2$ centered around $z = 0$, where the ion $i$ will fit according to its vdW radius.

\paragraph{Maximum velocity}
Since $v_{\rm b}$ is the maximum velocity between the amino acid and the longitudinal nanochannel surface as aforementioned, we can use Eq. \eqref{eq:vel_x} to obtain the max of $\hat{v}$, which is equivalent to $v_{\rm b}$. In this case we focus on the velocity within the bulk region, namely from the midpoint between the amino acid and the channel surface, $x = x_{\rm mid} = (R + r_{\circ}) / 2$, to the channel surface, $x = R$. We employ the following boundary conditions,
\begin{equation}
\begin{split}
&\hat{v}(x = R) = 0 , \\
&\hat{v}(x = x_{\rm mid} = \frac{R + r_{\circ}}{2}) \geq \hat{v}(x) ,
\end{split}
\label{eq:vel_x_bc}
\end{equation}
which are very similar to Eq. \eqref{eq:vel_bc}. By assuming a constant bulk concentration, $g_{\rm b}$, over this region we quickly come to a parabolic solution to Eq. \eqref{eq:vel_x} as well as determining $v_{\rm b} = q \tilde{z} g_{\rm b} E_{\perp} (R - r_{\circ})^2 / 4 \mu$, where $\tilde{z}_i$ has been simplified to $\tilde{z}$ since both ion species have the same valency. Then we have $v_{\rm b} = 154.47$ m/s by choosing a reasonable $r_{\circ} = 4$ {\AA}, a necessity in making $v_{\rm b}$ independent of the amino acid under study, which is more likely in experiment. In fact, the absolute value of the velocity determined from the Stokes equation is known to differ from experiment, \cite{qiao2003ion}, as opposed to the velocity ratio that we have utilized thus far. However, these differences appear to be systematic, \cite{qiao2003ion}, and can be solved by dividing $v_{\rm b}$ in half, resulting in the corrected $v_{\rm b} = 77.23$ m/s.
\clearpage
\renewcommand\thefigure{S\arabic{figure}}
\setcounter{figure}{0}
\renewcommand\thetable{S\arabic{table}}
\setcounter{table}{0}
\begin{table*}[t]
\caption{\label{table:eff_rad}Effective radii in {\AA}}
\begin{center}
\begin{tabular}{ c | c | c }
Amino & $r_{\rm eff}$ & $r_{\rm eff}$ \\
Acid & for ${\rm Cl}^-$ & for ${\rm K}^+$ \\
\hline ALA & 8.28 & 7.68 \\
\hline ARG & 3.50 & 4.49 \\
\hline ASN & 5.23 & 5.55 \\
\hline ASP & 6.15 & 4.69 \\
\hline CYS & 7.15 & 6.85 \\
\hline GLN & 4.85 & 5.28 \\
\hline GLU & 8.03 & 7.44 \\
\hline GLY & 6.18 & 5.75 \\
\hline HIS & 6.30 & 6.08 \\
\hline ILE & 5.25 & 4.78 \\
\hline LEU & 5.97 & 5.74 \\
\hline LYS & 5.38 & 6.43 \\
\hline MET & 4.92 & 4.73 \\
\hline PHE & 7.87 & 7.84 \\
\hline PRO & 5.02 & 4.43 \\
\hline SER & 6.85 & 6.99 \\
\hline THR & 4.15 & 4.57 \\
\hline TRP & 6.59 & 6.74 \\
\hline TYR & 6.93 & 7.29 \\
\hline VAL & 5.37 & 5.27
\end{tabular}
\end{center}
\end{table*}
\end{document}